\documentstyle[11pt,epsf]{article}
\begin{document}

\title{ASTROPHYSICAL IMPLICATIONS OF\\
NEUTRINO MASS AND MIXINGS
\footnote{To be published in the Proceedings of the Conerence on
Intersections Between Particle and Nuclear Physics, St. Petersburg,
FL, 1994}\\}

\author{ A.B. Balantekin \\
Physics Department, University of Wisconsin\\
Madison, WI 53706 }
\date{}

\maketitle

\begin{abstract}

Astrophysical implications of neutrino mass and mixings are
discussed. The status of solar and atmospheric neutrino problems, and
recent developments concerning nuclear physics
input to solar models and solar opacities are reviewed. Implications
of neutrino mass and mixings in supernova dynamics are explored.
The effects of supernova density fluctuations in neutrino propagation
is described.
\end{abstract}

\section*{INTRODUCTION}

\indent

Neutrinos, Pauli's little neutral ones, have already reached adolescence, and
they may eventually be regarded as the most elucidative indicators of new
astrophysical phenomena. In this review, some of the astrophysical
implications of neutrino mass and mixings are discussed.

Only certain values of neutrino mass and mixings give
rise to interesting astrophysical effects.
One should raise the question if these values are realistic. Unfortunately we
still do not have much experimental information about neutrino
properties. The only {\it measured} neutrino property is the number of
light neutrino flavors. It is determined from the invisible width of Z to be
\cite{par}

\begin{equation}
N_{\nu} = 2.99 \pm 0.04.
\end{equation}

\noindent
This number is consistent with the primordial nucleosynthesis limit from the
observed He abundance \cite{schramm}. We have {\it upper bounds} for all the
other neutrino properties : masses, flavor mixing angles, electromagnetic
moments, charges, and charge radii. In particular, the mass of the heaviest
neutrino, $\nu_{\tau}$, is very poorly known: upper limit is 35 MeV. A
tau neutrino
with a mass in the range of 1 to 10 eV could possibly be a significant
component of dark matter and, as we shall see later, could also have very
interesting implications for supernova dynamics.

In the next two sections first the status of solar neutrino problem is briefly
reviewed, and then some recent developments concerning physics input to solar
models are discussed. After reviewing the status of the atmospheric neutrino
anomaly, implications of neutrino mass and mixings in supernova dynamics are
explored. Finally neutrino propagation in inhomogeneous media is
described and its implications for the sun and supernovae are considered.

\section*{SOLAR NEUTRINOS}

\indent

Some of the energy released in the thermonuclear reactions in
the solar interior is emitted in the form of neutrinos and this
neutrino flux can be
calculated with relatively high precision \cite{bahc,baha}. This flux was
measured \cite{cl} and its directionality (i.e. coming from the sun) was
established \cite{kam}. However, the observed solar neutrino flux is deficient
relative to what is predicted by the standard solar model. A summary of the
current status of the solar neutrino experiments is given in Table 1.
Since the same thermonuclear reactions produce photons as well as neutrinos,
assuming the dominance of the $pp$ reaction, one can roughly estimate the
flux of the neutrinos coming from this reaction from the solar photon
luminosity to be

\begin{equation}
\phi_{\nu} \sim {2 L_\odot \over 26.73 {\rm MeV} - 2 {\overline E}_{\nu} }
{1\over 4 \pi r^2} \sim 6.5 \times 10^{10} cm^{-2} - s^{-1},
\end{equation}

\noindent
which indicates a Ga capture rate greater than 78 SNU's. Inclusion of
$^7$Be and $^8$B neutrinos would increase this number.

\begin{table}[b]
\vspace{8pt}
\centering
\begin{tabular}{|c|c|c|c|}
\hline
{\em Experiment} & {\em Data} & { SSM1 \cite{ssmb}} &{SSM2
\cite{ssmtc}} \\ \hline
 Homestake\cite{cl}  &  $2.3 \pm 0.3$ SNU & $8.0 \pm 1.0$ SNU & $6.4 \pm 1.4$
SNU \\ \hline
Kamioka \cite{kam}  &  $0.50 \pm 0.04 \pm 0.06$ & 1 & 0.78\\ \hline
SAGE \cite{sage1,sage2}  &  $74 ^{+13}_{-12}$ (stat.)$^{+5}_{-7}$ (sys.) SNU
& $131.5 ^{+7}_{-6}$ SNU & $122.5 \pm 7 $ SNU\\ \hline
Gallex \cite{gallex1,gallex2}  &  $ 79 \pm 10 $ (stat.) $\pm 6$ (syst.) SNU
& $131.5 ^{+7}_{-6}$ SNU & $122.5 \pm 7 $ SNU\\ \hline
\end{tabular}
\caption{The present status of solar neutrino experiments. The standard solar
models SSM1 and SSM2 are taken from Refs. [11] and [12]
respectively. The Gallex
and SAGE quotes are combined results of initial and more recent runs.}
\vspace{8pt}
\end{table}

The data from these experiments appear to be consistent with a stronger
suppression of the intermediate-energy neutrinos ($^7$Be) than the lower
energy ($pp$) or higher energy ($^8$B) neutrinos. It was recognized that, if
neutrinos observed in physical processes are mixtures of mass eigenstates,
coherent forward scattering of neutrinos in electronic matter gives rise to a
density-dependent effective mass resulting in an almost complete conversion of
electron neutrinos into neutrinos of a different flavor. This scenario, dubbed
the Mikheyev-Smirnov-Wolfenstein (MSW) effect, does not require fine-tuning of
mass differences and mixing angles and could possibly provide an elegant
solution to the solar neutrino problem within current theoretical prejudices
\cite{msw,others}. The non-adiabatic MSW solution with $\delta m^2 \sim
10^{-5}$ eV$^2$ and $\sin \theta \sim 0.1$ is consistent  with the data from
those four experiments. This MSW solution is still plausible after
incorporating the uncertainties in the solar model \cite{hata}. An
astrophysical
solution for the solar neutrino problem is unlikely, but still not
ruled out \cite{hata,scilla1}.

Solar neutrinos are not the only experimental probes of the sun. Information
from helioseismological p-wave observations complement information obtained by
solar neutrino experiments. The p-waves cause the solar surface to
vibrate with a characteristic period of about five minutes. By
observing red- and blue-shifts of
patches of the solar surface, projecting them on spherical harmonics, and
finally Fourier transforming with respect to the observation time one can
obtain eigenfrequencies of the solar p-modes. (One should exercise a little
bit of caution with regard to using spherical harmonics, since we only observe
half the sun). For very high overtones (for a spherically-symmetric
three-dimensional object such as the sun these are characterized by two
large integers), the equations describing p-modes simplify \cite{who}
and one can reliably obtain a
sound velocity profile for the outer half of the sun. The sound
density profile obtained this way agrees with the
predictions of the standard solar model. By studying
discontinuities in the sound velocity profile, it is also possible to reliably
extract the location of the bottom of the convective zone \cite{chris}.

\section*{PHYSICS INPUT IN SOLAR MODELS}

\indent

Physics input into solar models is extensively discussed in the literature,
see for example Ref. \cite{baha}. Here I concentrate on nuclear physics input
and the opacity.

There are two places where input from nuclear physics is needed for
solar neutrino
studies : i) to solar models, such as the rates of the reactions
leading to neutrinos, ii) to the detector cross sections, especially
cross sections for chemical detectors, such as chlorine, gallium, and
iodine. The
latter kind of input is not trivial to obtain, for example the utility of
(p,n) reactions to extract Gamow-Teller strengths was criticized \cite{eric}.
Here I will discuss the former kind of input, since some recent measurements
raised questions about previously-used values of S$_{17}$.

A measurement at RIKEN of the Coulomb breakup cross section of $^8$B
by $^{208}$Pb into p and $^7$Be was used to extract the cross section
for the reaction $^8$B
$+ \gamma \rightarrow ^7$Be + p, from which one can obtain the cross section
for the inverse reaction \cite{riken}. The resulting astrophysical S-factor
S$_{17}$ is less than those obtained from previous measurements which
would reduce the predicted flux of $^8$B neutrinos. The reaction
measured at RIKEN is an
electromagnetic (Coulomb breakup) process. In this experiment, to eliminate
contributions from the strong nuclear force data are taken at very forward
angles (i.e., at large impact parameters). Hence only the $E1$ contribution is
measured. Caution should be exercised when extracting S$_{17}$ at low
energies since other multipoles, not measured at RIKEN, are expected
to contribute to it as well.
There are sources of uncertainties in the method of virtual
quanta used here (such as determining the value of the lowest impact parameter)
\cite{carlos}. It is probably too early to tell if this experiment warrants a
lower S$_{17}$ before more data are taken.

Opacity is another parameter in solar models which had to be recently modified
\cite{rog}. In the radiative zone of the sun, heat transfer by radiation is
controlled by a single opacity parameter, which is a measure of
photoabsorption. To calculate opacity one needs to include photoabsorption
cross sections for many types of atomic configurations; hence in principle,
one needs to know all atomic levels of all the isotopes. Heavier elements
(those with a higher Z) contribute more to the opacity. For example iron
alone, which is basically a trace element in the sun, contributes about 20\%.
Changing the opacity alters the rate of heat transfer and consequently the
core temperature. By lowering the opacity one can homologously lower
\cite{scilla} the temperature profile. Since the rates of the
neutrino-producing subbarrier fusion reactions in the sun are governed by
quantum-mechanical tunneling, lowering the temperature lowers the rate almost
exponentially. The flux of higher-energy neutrinos coming from reactions with
higher Coulomb barriers rapidly decreases as the core temperature falls.
(This exponential dependence is usually expressed
as a power law for a limited range of temperatures near the standard
solar model temperature. This practice can sometimes be misleading.)

Until recently, most researchers used the Los Alamos Opacity Library
\cite{laol}. However a number of problems with pulsating stars
indicate a need for
increasing these opacities in the convective zone \cite{stone}. It was also
recently
shown that for agreement between the results of the standard solar model and
the helioseismologically observed p-mode frequencies, the solar opacity needs
to be increased by about 30\% from the Los Alamos Opacity Library
results \cite{guenther}. Indeed, recent calculations at Livermore increase the
opacity by about 15\% in the convective zone and about 5\% in the core
\cite{rog}. Increasing the opacity, however, raises the core temperature, and
consequently increases the neutrino flux and could eliminate the effect of a
lowered S$_{17}$.

\section*{ATMOSPHERIC NEUTRINOS}

\indent

Atmospheric neutrinos arise from the decay of secondary pions, kaons, and
muons produced by the collisions of primary cosmic rays with the $O$ and $N$
nuclei in the upper atmosphere. For energies less than 1 GeV all the
secondaries decay :

\begin{eqnarray}
\pi^{\pm} (K^{\pm}) &\rightarrow &\mu^{\pm} + \nu_{\mu} (\overline{\nu}_{\mu}),
\nonumber\\
\mu^{\pm} &\rightarrow & e^{\pm} + \nu_e (\overline{\nu}_e) +
\overline{\nu}_{\mu} (\nu_{\mu}).
\end{eqnarray}

\noindent
Consequently one expects the ratio

\begin{equation}
r = (\nu_e + \overline{\nu}_e) / (\nu_{\mu} + \overline{\nu}_{\mu})
\end{equation}

\noindent
to be approximately 0.5 in this energy range. Detailed Monte Carlo
calculations \cite{gaisser}, including the effects of muon polarization, give $
r \sim 0.45$. Since one is evaluating a ratio of similarly calculated
processes, systematic errors are significantly reduced. Different
groups estimating
this ratio, even though they start with neutrino fluxes which can differ in
magnitude by up to 25\%, all agree within a few percent
\cite{bludman}. The ratio (observed to predicted) of ratios

\begin{equation}
R = {(\nu_{\mu} / \nu_e)_{\rm data} \over (\nu_{\mu} / \nu_e)_{\rm Monte
Carlo} }
\end{equation}

\noindent
was determined in several experiments as summarized in Table 2. There seems to
be a persistent discrepancy between theory and experiment. Oscillations
of $\nu_{\mu}$ into $\nu_{\tau}$ are generally invoked to explain this
discrepancy \cite{osc}.

\begin{table}[b]
\vspace{8pt}
\centering
\begin{tabular}{|c|c|}
\hline
{\em Experiment} & {\em R} \\ \hline
\indent Kamioka \cite{kami}  &  $0.60 ^{+0.07}_{-0.06} \pm 0.05$ \\ \hline
\indent IMB \cite{imb}  &  $0.54 \pm 0.05 \pm 0.12$ \\ \hline
\indent Soudan \cite{soudan}  & $0.55 \pm 0.27$ \\ \hline
\indent Nusex \cite{nusex}  &  $0.99 ^{+0.35}_{-0.25}$\\ \hline
\indent Frejus \cite{frejus}  &  $0.87 \pm 0.16 \pm 0.08$ \\ \hline
\end{tabular}
\caption{
Ratio of Ratios, R of Eq. (5), as observed in different experiments.}
\vspace{8pt}
\end{table}

Experimentally the ratio of ratios, $R$, appears to be independent of zenith
angle. The observed zenith angle distribution of low energy atmospheric
neutrinos is consistent with no oscillations or with a large number of
oscillations for all source-detector distances. Explanations of the low energy
atmospheric neutrino anomaly based on the oscillation of 2 neutrino flavors
require that the oscillating term \cite{bilen}, $\cos(\frac{\delta m^2 L}{2
E})$, average to zero for even the shortest source-detector distances ($L< 50$
km for neutrinos from directly overhead.) For neutrinos in the energy range
0.1 to 1 GeV this condition is satisfied for $\delta m^2 > 10 ^{-3} eV^2$.
If the atmospheric neutrino anomaly is resolved by the oscillations of
muon into tau neutrinos, this value of $\delta m^2$ is consistent with
a tau neutrino mass relevant to hot dark matter and supernova
dynamics.
It is also possible to make a search in the three-neutrino-flavor parameter
space and identify regions in this parameter space compatible with the
existing atmospheric and solar neutrino data within the vacuum oscillation
scheme \cite{andy}.

\section*{NEUTRINO FLAVOR MIXING IN SUPERNOVAE}

\indent

Understanding neutrino transport in a supernova is an essential part of
understanding supernova dynamics. In this review, I will only
concentrate on the
effects of neutrino flavor mixing on supernova dynamics.
\begin{figure}[h]
\vspace{8pt}
\centerline{\hbox{\epsfxsize= 4in \epsfbox[105 242 459 518]{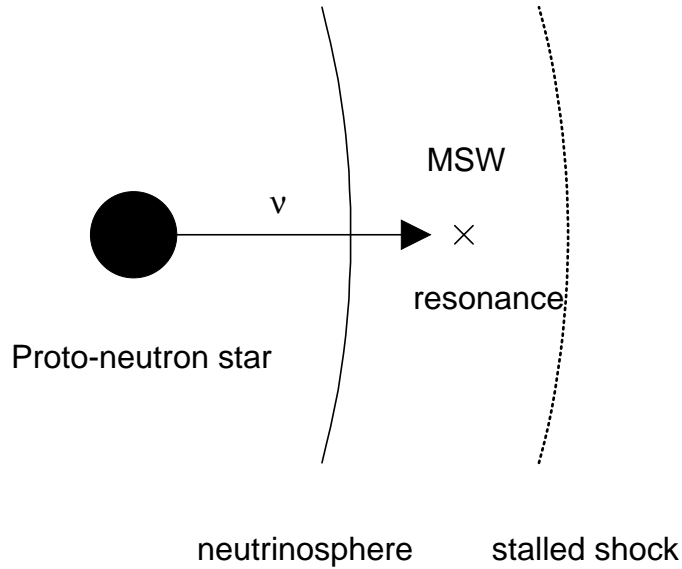}}}
\caption{
Neutrinosphere and the stalled shock in a core-collapse driven supernova.}
\vspace{8pt}
\end{figure}
In a core-collapse
driven supernova, the inner core collapses subsonically, but the outer part of
the core supersonically. At some point during the collapse, when the nuclear
equation of state stiffens, the inner part of the core bounces, but the outer
core continues falling in. The shock wave generated at the boundary loses its
energy as it expands by dissociating material falling through it into free
nucleons and alpha particles. For a large initial core mass, the shock wave
gets stalled at $\sim$ 200 to 500 km away from the center of the proto-neutron
star \cite{mayle}. Meanwhile, the proto-neutron star, shrinking under its own
gravity, loses energy by emitting neutrinos, which only interact weakly and
can leak out on a relatively long diffusion time scale. The question to
be investigated then is
the possibility of regenerating the shock by neutrino heating.

The situation at the onset of neutrino heating is depicted in Figure 1. The
density at the neutrinosphere is $\sim 10^{12}$g cm$^{-3}$ and the density at
the position of the stalled shock is $\sim 2 \times 10^7$ g cm$^{-3}$
\cite{mayle}. Writing the MSW resonance density in appropriate units :

\begin{equation}
\rho_{\rm res} = 1.31 \times 10^7 \left( {\delta m^2 \over {\rm eV}^2} \right)
\left( { {\rm MeV} \over E_{\nu} } \right) \left( {0.5 \over {\rm Y}_e}
\right) {\rm g} \> {\rm cm}^{-3},
\end{equation}

\noindent
one sees that, for small mixing angles, $E_{\nu} \sim 10$ MeV, and
cosmologically interesting $\delta m^2 \sim 1 - 10^4$ eV$^2$, there is an MSW
resonance point between the neutrinosphere and the stalled shock.

Neutrinos emitted from the core are produced by a neutral current process,
and so the luminosities are approximately the same for all flavors.
The energy spectra are
approximately Fermi-Dirac with a zero chemical potential characterized by a
neutrinosphere temperature. The $\nu_{\tau}, {\overline
\nu}_{\tau}, \nu_{\mu}, {\overline \nu}_{\mu}$ interact with matter only via
neutral current interactions. These decouple at relatively small radius and end
up with somewhat high temperatures, about 8 MeV. The ${\overline \nu}_e$'s
decouple at a larger radius because of the additional charged current
interactions with the protons, and consequently have a somewhat lower
temperature, about 5 MeV. Finally, since they undergo charged current
interactions with more abundant neutrons, $\nu_e$'s decouple at the largest
radius and end up with the lowest temperature, about 3.5 to 4 MeV.
An MSW resonance between the neutrinosphere and the stalled shock can then
transform $\nu_{\tau} \leftrightarrow \nu_e$, cooling $\nu_{\tau}$'s, but
heating $\nu_e$'s. Since the interaction cross section of
electron neutrinos with the matter in the stalled shock increases with
increasing energy, it may be
possible to regenerate the shock. Fuller {\it et al.} found that for small
mixing angles between $\nu_{\tau}$ and $\nu_e$ one can get a 60\% increase
in the explosion energy \cite{mayle}.

There is another implication of the $\nu_{\tau}$ and $\nu_e$ mixing in the
supernovae. Supernovae are possible r-process sites \cite{burb}, which requires
a neutron-rich environment, i.e., the ratio of electrons to baryons, $Y_e$,
should be less than one half. $Y_e$ in the nucleosynthesis region is given
approximately by \cite{qian}

\begin{equation}
Y_e \simeq {1 \over 1+ \lambda_{{\overline \nu}_e p} / \lambda_{ \nu_e n}}
\simeq {1 \over 1 + T_{{\overline \nu}_e} / T_{ \nu_e}},
\end{equation}

\noindent
where $\lambda_{ \nu_e n}$, etc. are the capture rates. Hence if $T_{{\overline
\nu}_e} > T_{\nu_e}$, then the medium is neutron-rich. As we discussed
above, without matter-enhanced neutrino oscillations, the neutrino
temperatures satisfy the inequality $T_{ \nu_{\tau}} >T_{{\overline \nu}_e} >
T_{ \nu_e}$. But the MSW effect, by heating $\nu_e$ and cooling $\nu_{\tau}$
can reverse the direction of inequality, making the medium proton-rich
instead. Hence the existence of neutrino mass and mixings puts severe
constraints on heavy-element nucleosynthesis in supernova. These constraints
are investigated in Ref. \cite{qian}.

\section*{EFFECTS OF DENSITY FLUCTUATIONS}

\indent

If a completely polarized particle beam travels through a random magnetic
field for a long enough time, it will be completely depolarized. In a similar
way, if different neutrino flavors mix, a completely ``polarized''
neutrino beam
(say all $\nu_e$) may become completely ``depolarized'' (half $\nu_e$, half
$\nu_x$) after passing through a medium with fluctuating matter density
\cite{frank1}. Note that this can happen without a neutrino magnetic moment,
it is simply a new aspect of the MSW mechanism.
There is an extensive discussion of neutrino oscillations in
inhomogeneous media in the literature \cite{frank1,wick}.
\begin{figure}[h]
\vspace{8pt}
\centerline{\hbox{\epsfxsize= 4in \epsfbox[77 53 538 497]{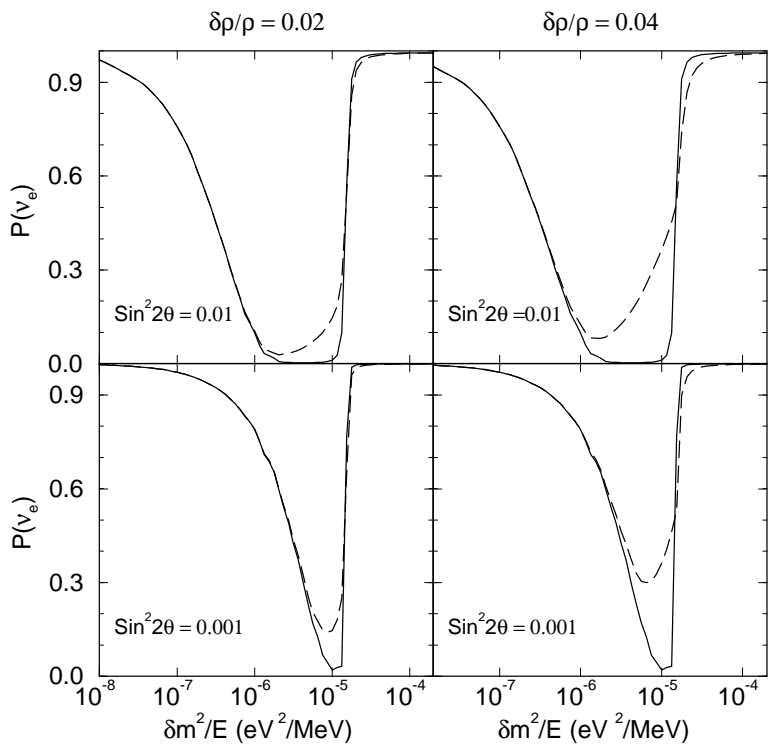}}}
\caption{
MSW effect in the sun with (dashed lines) and without (solid lines) density
fluctuations.}
\vspace{8pt}
\end{figure}

We take the total electron density to be $\rho_e (r) + \delta \rho_e (r)$
where $\delta \rho_e (r)$ is the fluctuating part. One can assume that the
average fluctuation vanishes

\begin{equation}
< \delta \rho_e (r) >  = 0,
\end{equation}

\noindent
but the two-body correlations are non-zero :

\begin{equation}
< \delta \rho_e (r) \delta \rho_e (r') >  = \rho_0^2 f(|r-r'|),
\end{equation}

\noindent
where the correlation function $f(|r-r'|)$ has a finite correlation length. The
effect is significant if the correlation length, $r_c$, is small
\cite{frank1}.

In the white noise limit,

\begin{equation}
f(|r-r'|) \rightarrow 2 r_c \delta (r-r'),
\end{equation}

\noindent
a number of simplifications make numerical calculations particularly easy.
The effects of possible solar density fluctuations are presented in Figure 2.
In this figure the electron neutrino survival probability is plotted for two
values of the mixing angle, chosen to provide non-adiabatic solutions for 2\%
(left column) and 4\% density fluctuations (right column). The density
fluctuations are taken to be proportional to the local density predicted by
the standard solar model \cite{ssmb}. One observes that the effect of density
fluctuations in the sun are small, but can conceivably give rise to annual
variations in the solar neutrino flux.

The MSW mechanism may also be relevant in collapsing
pre-supernova stellar cores,
where adiabatic conversion of electron neutrinos into massive (e.g.
$\nu_{\tau}$) neutrinos could result in readjustment of lepton numbers and
small entropy generation \cite{fuller}. A consequent drop in the electron
fraction, $Y_e$, could be significant for the mechanism of supernova
explosion.

\begin{figure}[t]
\vspace{8pt}
\centerline{\hbox{\epsfxsize= 4in \epsfbox[36 53 558 557]{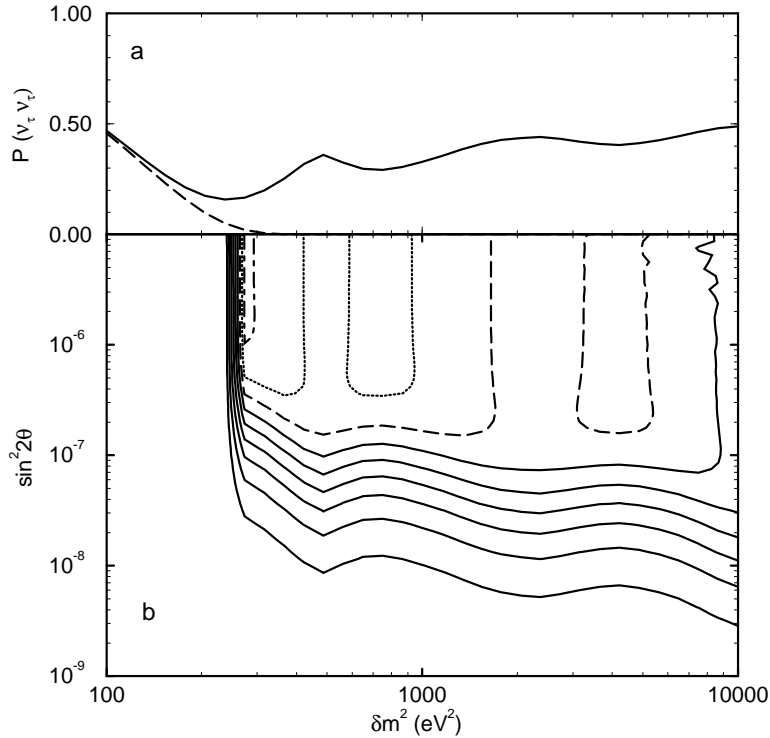}}}
\caption{The effect of density fluctuations at the infall [42]. a)
Neutrino survival probability with (solid line) and without (dashed line)
fluctuations. b) $\nu_e \rightarrow \nu_{\tau}$ transition probability
contours (see text).
}
\vspace{8pt}
\end{figure}

The effect of density fluctuations in a collapsing star is plotted in Figure
3. In this figure the upper panel exhibits the neutrino survival
probability as a function of the mass difference squared between the electron
and heavy neutrino with (solid line) and
without (dashed line) density fluctuations. The fractional density
fluctuations are taken to be $5 \times 10^{-3}$. One observes that there is a
complete depolarization (50\% conversion), especially for heavy neutrino mass
(assuming a very small electron neutrino mass) greater than $\sim 30$ MeV.
In the lower panel the $\nu_e \rightarrow \nu_{\tau}$ transition probability
contours are plotted. Here the lowest solid line indicates 10\%,
dashed lines 70\%, and the dotted lines 80\% conversion, with intermediate
lines being in steps of 10\%. For a detailed study of density fluctuations in
supernova, both during the infall and at the hot-bubble after the bounce, the
reader is referred to Ref. \cite{frank2}.

\section*{CONCLUSIONS}

\indent

Neutrino astrophysics is a field with considerable prospects. In this
review I only discuss low energy neutrino sources. There are
many physics questions, such as the nature of the central engines of
active galactic nuclei,  which can be explored by doing high energy
astrophysics \cite{halzen}. Most of the time the information obtained from
low- and high-energy neutrino astrophysics is complementary. For example,
it is possible to look for high-energy neutrino signatures of cold dark matter.
Supernova neutrino observations and long-baseline neutrino
experiments at somewhat lower energies
probe hot dark matter. These experiments together may help us
assess the role of the dark
matter in the structure of the universe. I believe that in the years
to come, neutrino astrophysics will answer many fundamental questions
and will pose many new ones.

\section*{ACKNOWLEDGEMENTS}

\indent

This research was supported in part by the U.S. National Science Foundation
Grant No. PHY-9314131, and in part by the University of Wisconsin Research
Committee with funds granted by the Wisconsin Alumni Research Foundation.


\begin{thebibliography}{10}

\bibitem{par}  Particle Data Group, {\it Review of Particle Properties}, 1992
Edition.
\bibitem{schramm} J. Yang, M. Turner, G. Steigman, D. Schramm, and K. Olive,
Astrophys. J. {\bf 281}, 493 (1984).
\bibitem{bahc} J. N. Bahcall, {\it Neutrino Astrophysics}, (Cambridge
University Press, Cambridge, 1989).
\bibitem{baha} {\it Solar Modelling}, A.B. Balantekin and J.N. Bahcall, Eds.,
(World Scientific, Singapore, 1994).
\bibitem{cl}
R. Davis, D. S. Harmer, and K. C. Hoffman, Phys. Rev. Lett. {\bf20}, 1205
(1968); J. K. Rowley, B. T. Cleveland, and R. Davis in {\it Solar Neutrinos
and Neutrino Astronomy (Lead High School, South Dakota)}, Proceedings,
edited by M. L. Cherry, K. Lande, and W. A. Fowler, AIP Conf. Proc. No. 126
(AIP, New York, 1985), p. 1.
\bibitem{kam}
K. S. Hirata {\it et al.}, Phys. Rev. Lett. {\bf 63}, 16 (1989); {\bf65},
1297 (1990);{\bf65}, 1301 (1990); K. Nakamura, to be published (1994).
\bibitem{sage1}
A.I. Abazov {\it et al.}, Phys. Lett. B {\bf 328}, 234 (1994); Phys. Rev.
Lett. {\bf 67}, 3332 (1991).
\bibitem{sage2}
A.I. Abazov {\it et al.}, these proceedings.
\bibitem{gallex1}
P. Anselmann {\it et al.}, Phys. Lett. B {\bf 285}, 190 (1992).
\bibitem{gallex2}
P. Anselmann {\it et al.}, Phys. Lett. B {\bf 327}, 377 (1994).
\bibitem{ssmb}
J. N. Bahcall, W. F. Huebner, S. H. Lubow, P. D. Parker, and R. K. Ulrich,
Rev. Mod. Phys. {\bf 54}, 767 (1982); J.N. Bahcall and R. Ulrich, Rev. Mod.
Phys. {\bf 60}, 297 (1988); J.N. Bahcall and M.H. Pinsonneault, Rev. Mod.
Phys. {\bf 64}, 885 (1992).
\bibitem{ssmtc}
S. Turck-Chieze and I. Lopes, Astrophys. J. {\bf 408}, 347 (1993).
\bibitem{msw}
L. Wolfenstein, Phys. Rev. D {\bf 17}, 2369 (1978); {\bf 20}, 2634
(1979); S.P. Mikheyev and A. Yu. Smirnov, Nuovo Cim. {\bf 9C}, 17 (1986); Sov.
J. Nucl. Phys. {\bf 42}, 913 (1986).
\bibitem{others}
H.A. Bethe, Phys. Rev. Lett. {\bf 56}, 1305 (1986); S.P. Rosen and J.M.
Gelb, Phys. Rev. D {\bf 34}, 969 (1986); E.W. Kolb, M.S. Turner, and T.P.
Walker, Phys. Lett. B {\bf 175}, 478 (1986); W.C. Haxton, Phys. Rev. Lett.
{\bf 57}, 1271 (1986); Phys. Rev. D {\bf 35}, 2352 (1987); V. Barger, K.
Whisnant, S. Pakvasa, and R.J.N. Phillips, {\it ibid.} {\bf 22}, 2718 (1980);
V. Barger, R.J.N. Phillips, and K. Whisnant, {\it ibid.} {\bf 34}, 980 (1986);
S.J. Parke, Phys. Rev. Lett. {\bf 57}, 1275 (1986); T.K. Kuo and J.
Pantaleone, Phys. Rev. D {\bf 35}, 3432 (1987); C.W. Kim, S. Nussinov, and W.K.
Sze, Phys. Lett. B {\bf 184}, 403 (1987); A. Dar, A. Mann, Y. Melina, and D.
Zajfman, Phys. Rev D {\bf 35}, 3607 (1987); A.B. Balantekin, S.H. Fricke,
and P.J. Hatchell, {\it ibid.} {\bf 38}, 935 (1988); A.J. Baltz, and J.
Weneser, {\it ibid.} {\bf 37}, 3364 (1988).
\bibitem{hata}
N. Hata, in Ref. \cite{baha}.
\bibitem{scilla1}
V. Castellani, S. Degl'Innocenti, G. Fiorentini, M. Lissia, and
B. Ricci, Phys. Lett. B {\bf 324}, 425 (1994).
\bibitem{who}
D. Gough, Solar Phys. {\bf 100}, 65 (1985).
\bibitem{chris}
J. Christensen-Dalsgaard, D.O. Gough, and M.J. Thompson, Astrophys. J. {\bf
378}, 413 (1991).
\bibitem{eric}
A. Garcia {\it et al.}, Phys. Rev. Lett. {\bf 67}, 3654 (1991).
\bibitem{riken}
T. Motobayashi {\it et al.}, RIKEN preprint Rikkyo RUP 94-2; see also K.
Langanke and T.D. Shoppa, Phys. Rev. C {\bf 49}, R1771 (1994).
\bibitem{carlos}
G. Baur, C.A. Bertulani, and H. Rebel, Nucl. Phys. A {\bf 458}, 188 (1986).
\bibitem{rog}
F. Rogers and C. Iglesias, Science {\bf 263}, 50 (1994).
\bibitem{scilla}
V. Castellani, S. Degl'Innocenti, G. Fiorentini, and B. Ricci, in Ref.
\cite{baha}.
\bibitem{laol}
W.F. Huebner, A.L. Merts, N. H. Magee, M.F. Argo, Los Alamos Sci. Rep.
La-6760-M (1977).
\bibitem{stone}
J.O. Peterson, Astron. Astrophys. {\bf 265}, 555 (1992).
\bibitem{guenther}
D.B. Guenther, P. Demarque, Y.-C. Kim, and M.H. Pinsonneault, Astrophys. J.
{\bf 387}, 377 (1982).
\bibitem{gaisser}
G. Barr, T.K. Gaisser, and T. Stanev, Phys. Rev. D {\bf 39}, 3532 (1989);
T.K. Gaisser, T. Stanev, and G. Barr, {\it ibid.} {\bf 38}, 85 (1988).
\bibitem{bludman}
H. Lee and S.A. Bludman, Phys. Rev. D {\bf 37}, 122 (1988);
E.V. Bugaev and V.A. Naumov, Phys. Lett. B {\bf 232}, 391 (1989);
M. Honda, K, Kasahara, K. Hidaka, and S. Midorikawa, Phys. Lett. B {\bf 248},
193 (1990);
M. Kawasaki and S. Mizuta, Phys. Rev. D {\bf 43}, 2900 (1991).
\bibitem{kami}
K.S. Hirata {\it et al.}, Phys. Lett. B {\bf 205}, 416 (1988).
\bibitem{imb}
D. Casper {\it et al.}, Phys. Rev. Lett. {\bf 66}, 2561 (1991).
\bibitem{soudan}
P. Litchfield (Soudan 2 Collaboration), Proc. Int. Workshop on $(\nu_{\mu} /
\nu_e)$ Problem in Atmospheric Neutrinos, Gran Sasso, Italy, March 1993.
\bibitem{nusex}
M. Aglietta {\it et al.}, Europhys. Lett. {\bf 8}, 611 {1989}.
\bibitem{frejus}
C. Berger {\it et al.}, Phys. Lett. B {\bf 245}, 305 (1989).
\bibitem{osc}
J.G. Learned, S. Pakvasa, and T. Weiler, Phys. Lett. B {\bf 207}, 79 (1988);
V. Barger and K. Whisnant, {\it ibid.} {\bf 209}, 365 (1988); K. Hidaka, M.
Honda, and S. Midorikawa, Phys. Rev. Lett. {\bf 61}, 1537 (1988).
\bibitem{bilen}
S.M. Bilenky and S.T. Petcov, Rev. Mod. Phys. {\bf 59}, 671 (1987).
\bibitem{andy}
A. Acker, A.B. Balantekin, and F.N. Loreti, Phys. Rev. D {\bf 49}, 328 (1994).
\bibitem{mayle}
G.M. Fuller, R.W. Mayle, B.S. Meyer, and J.R. Wilson, Astrophys.
J. {\bf 389}, 517 (1992);
R.W. Mayle, in {\it Supernovae}, A.G. Petschek, Ed., (Springer-Verlag, New
York, 1990).
\bibitem{burb}
E.M. Burbidge, G.R. Burbidge, W.A. Fowler, and F. Hoyle, Rev. Mod. Phys. {\bf
29}, 694 (1957).
\bibitem{qian}
Y.Z. Qian, G.M. Fuller, G.J. Mathews, R.W. Mayle, J.R. Wilson, and S.E.
Woosley, Phys. Rev. Lett. {\bf 71}, 1965 (1993).
\bibitem{frank1}
F.N. Loreti and A.B. Balantekin, Phys. Rev. D, submitted for publication
(1994).
\bibitem{wick}
W.C. Haxton and W.-M. Zhang, Phys. Rev. D {\bf 43}, 2484 (1991);
P.I. Krastev and A. Yu. Smirnov, Phys. Lett. B {\bf 226}, 341 (1989);
R.F. Sawyer, Phys. Rev. D {\bf 42}, 3908 (1990); A. Schafer and
S.E. Koonin, Phys. Lett. B {\bf 185}, 417 (1987).
\bibitem{frank2}
F.N. Loreti,Y.Z. Qian, A.B. Balantekin, and G.M. Fuller, in preparation.
\bibitem{fuller}
G.M. Fuller, R.W. Mayle, J.R. Wilson, and D.N. Schramm, Astrophys. J. {\bf
322}, 795 (1987).
\bibitem{halzen}
T.K. Gaisser, F. Halzen, and T. Stanev, Phys. Rep., in press.

\end{thebibliography}
\end{document}